\documentclass[%
superscriptaddress,
 amsmath,amssymb,
pra,
twocolumn,
]{revtex4-1}

\usepackage{graphicx}
\usepackage{dcolumn}
\usepackage{bm}
\usepackage{color}
\usepackage{hyperref}
\hypersetup{backref,colorlinks=true,breaklinks,urlcolor=blue,linkcolor=blue,citecolor=blue}
\usepackage{siunitx}
\usepackage{array}
\usepackage{booktabs}
\usepackage[table]{xcolor}
\usepackage{ulem}

\begin{document}

\title{Beating the classical precision limit with spin-1 Dicke state of more than 10000 atoms}

\author{Yi-Quan Zou}
\author{Ling-Na Wu}
\author{Qi Liu}
\author{Xin-Yu Luo}
\author{Shuai-Feng Guo}
\author{Jia-Hao Cao}
\affiliation{State Key Laboratory of Low Dimensional Quantum Physics, Department of Physics, Tsinghua University, Beijing 100084, China}
\author{Meng Khoon Tey}
\email{mengkhoon\_tey@mail.tsinghua.edu.cn}
\affiliation{State Key Laboratory of Low Dimensional Quantum Physics, Department of Physics, Tsinghua University, Beijing 100084, China}
\affiliation{Collaborative Innovation Center of Quantum Matter, Beijing, China}
\author{Li You}
\email{lyou@mail.tsinghua.edu.cn}
\affiliation{State Key Laboratory of Low Dimensional Quantum Physics, Department of Physics, Tsinghua University, Beijing 100084, China}
\affiliation{Collaborative Innovation Center of Quantum Matter, Beijing, China}


\date{\today}

\begin{abstract}
Interferometry is a paradigm for most precision measurements.
Using $N$ uncorrelated particles, the achievable
precision for a two-mode (two-path) interferometer is bounded by the standard quantum limit (SQL), $1/\sqrt{N}$,
due to the discrete (quanta) nature of individual measurements. Despite being a challenging benchmark, the two-mode SQL has been approached in a number of systems, including the LIGO and today's best atomic clocks. Employing multi-mode interferometry, the SQL becomes $1/[(M-1)\sqrt{N}]$ using M modes. Higher precision can also be achieved using entangled particles such that quantum noises from individual particles cancel out. In this work, we demonstrate an interferometric precision of $2.42^{+1.76}_{-1.29}\,$dB beyond the three-mode SQL, using balanced spin-1 (three-mode) Dicke states containing thousands of entangled atoms. The input quantum states are deterministically generated by controlled quantum phase transition and exhibit close to ideal quality. Our work shines light on the pursuit of quantum metrology beyond SQL.
\end{abstract}


\maketitle

Since introduced by Dicke in an effort to effectively explain superradiance
in 1954 \cite{Dicke1954PhysRev.93.99},
Dicke state has attracted widespread attention for its potential applications in quantum information and precision measurement \cite{giovannetti2004quantum,smerzi2016review}. For a collection of $N$ identical (pseudo-) spin-1/2 particles, Dicke states
map onto Fock states $|N/2+m\rangle_\uparrow |N/2-m\rangle_\downarrow$ with $(N/2+m)$ particles in spin-up $\uparrow$ and $(N/2-m)$
in spin-down $\downarrow$ modes for $m=-N/2,-N/2+1\cdots,N/2$.
The special case of one excitation
 $ |1\rangle_\uparrow|N-1\rangle_\downarrow$, or $\left( {| \uparrow  \downarrow \downarrow \cdots  \downarrow \rangle  + | \downarrow  \uparrow  \downarrow  \cdots  \downarrow \rangle  + \cdots + | \downarrow  \downarrow \cdots  \downarrow  \uparrow \rangle } \right)/\sqrt N $
in terms of the product state basis, is often called
 W state. It is potentially important for
quantum information due to its robustness to particle loss.
Another Dicke state of wide interest is the so-called twin-Fock state $|N/2\rangle_\uparrow |N/2\rangle_\downarrow$ (for even $N$). It has been demonstrated to allow measurement precision beyond the SQL \cite{Klempt2011ScienceTF, Burnett1993PhysRevLett.71.1355}, along with other entangled states such as squeezed light \cite{Roman2016PhysRevLett.117.110801}, squeezed spin state \cite{Wineland1992PhysRevA.46.R6797,Ueda1993PhysRevA.47.5138,Kasevich2016NatureQND,Thompson2016PhysRevLett.116.093602,Schmiedmayer2013integrated,Oberthaler2010Nature,riedel2010Nature,Bollinger2016science,MA201189}, and NOON state \cite{Dowling2002quantum,Blatt2011PhysRevLett.106.130506,Pan2016PhysRevLett.117.210502}.

Dicke states are not limited to ensemble of spin-$1/2$ particles. More generally, they are the common eigenstates $|l,m\rangle$ of the collective spin operators ${ \hat {\bf L}}^2$ and $\hat L_z$,
with respective eigenvalues $l(l+1)$ and $m$ ($\hbar=1$ hereafter). Here,
${ \hat { \bf L} }\equiv(\hat{L}_x,\hat{L}_y,\hat{L}_z)$, $\hat{L}_k = \sum\nolimits_{j=1}^N \hat{s}_k^{(j)}$ with $\hat{s}_k^{(j)}$ representing the
spin operator of the $j$-th particle
along the $k\,(=x,y,z)$ direction, applies for any spin $s$.
As Dicke state is an eigenstate of $\hat L_z$ ($=m$), the direction of its transverse spin is totally indeterminate
according to the Heisenberg uncertainty principle. Hence,
Dicke state can be represented as an annulus on the generalized Bloch sphere of radius $\sqrt{l(l+1)}$ (Fig. 1{\it A}).
Dicke states with $|m|\ne l$ constitute an important class of entangled states.
Tremendous progresses have been made at their generation over the past decades
 using photons \cite{Weinfurter2009PhysRevLett.103.020504,Zeilinger2009PhysRevLett.103.020503}, ions \cite{Blatt2005Nature} and cold atoms \cite{Reichel2014ScienceWstate,Klempt2011ScienceTF,Luo2017TwinFock}.
To our knowledge, all these generated Dicke states are based on pseudo-spin-1/2 particles so far, except for the heralded spin-1 W-state by detection of a single photon \cite{Vuletic2015NatLettWstate}.

This article reports the first generation of spin-1 Dicke states in the close vicinity of $|l=N,m=0\rangle$ with $N\approx11700$.
These states are deterministically generated by driving a condensate of spin-1 atoms through a quantum phase transition (QPT) \cite{zhangduan2013PRL}. Compared with our previously reported twin-Fock state \cite{Luo2017TwinFock} which makes use of only the $m_F=\pm1$ spin components of the atoms, the spin-1 Dicke state takes advantage of all three components and thus offers higher interferometric sensitivity \cite{Brand2012PhysRevLett.108.130402,Cataliotti2013NJP,Hirano2013JPS,Mitchell2012PhysRevLett.109.253605,Polzik2010PhysRevLett.104.133601,Smerzi2013PhysRevA.87.033607} ({\it Methods}). Using the prepared states, we demonstrate enhanced measurement precision beyond the SQL of three-mode interferometry.

\section*{Generation of spin-1 Dicke state through QPT}
In the absence of external electromagnetic fields and when the density-dependent spin-symmetric interaction dominates such that the same spatial wave function can be assumed for all spin components ($m_F=0,\pm1$), a spinor BEC in the ground hyperfine manifold $F=1$ is described by the Hamiltonian \cite{Law1998PhysRevLett.81.5257}
$H={c_2} {\hat {\bf L}}^2/(2N)$, with $c_2$ the spin-dependent interaction strength.
Here, $\hat{s}_k^{(j)}$ becomes the spin-1 operator $\hat{F}_k^{(j)}$ for
the $j$-th atom. The $^{87}$Rb spin-1 BEC
is ferromagnetic \cite{chapman2004PhysRevLett.92.140403} with $c_2<0$, its ground states thus correspond to the $(2N+1)$-fold degenerate Dicke states $|l=N,m\rangle$ which maximize ${\hat {\bf L}^2}$.
Among all, the Dicke state with the smallest $|m|$($=0$), or the balanced Dicke state, is the most entangled and it allows for the highest measurement precision ({\it Methods}).

\begin{figure}[t!]
\centering
\includegraphics[width=1\columnwidth]{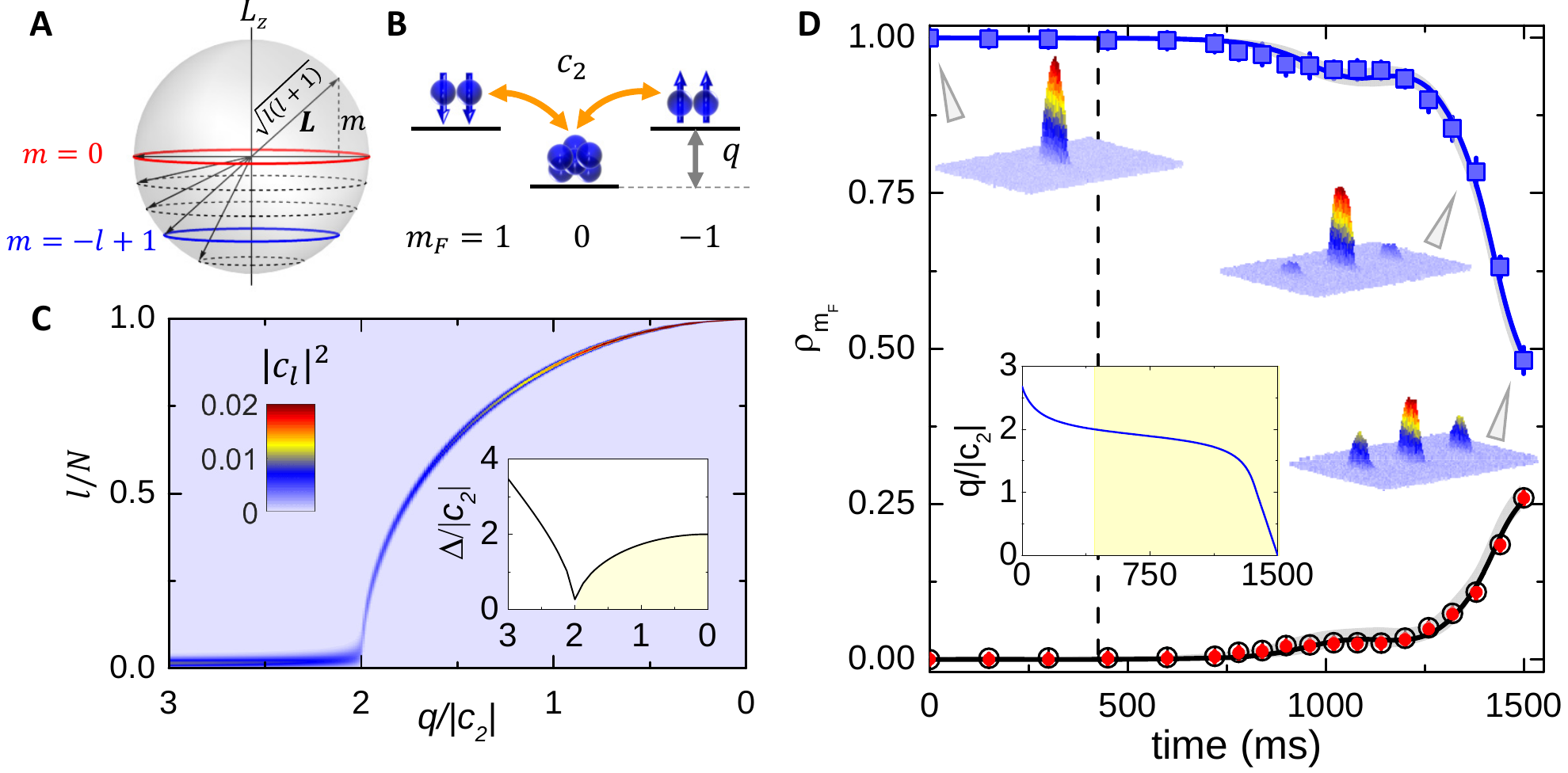}
\caption{{\bf Generation of spin-1 Dicke state.}
({\it A}) Dicke states represented on the generalized Bloch sphere of radius $\sqrt{l(l+1)}$ as annuli whose latitudes are determined by $m=-l,-l+1,..,l$.
The red solid annulus denotes the balanced Dicke state $|l,m=0\rangle$, and the blue solid annulus represents
$|l,m=-l+1\rangle$, ($l=4$ here).
 ({\it B}) A $^{87}$Rb atomic BEC in the $F=1$ hyperfine state undergoing spin-mixing dynamics at rate $\propto|c_2|$,
 in the presence of a quadratic Zeeman shift $q$. The linear Zeeman shifts are not shown.
 ({\it C}) The distribution of projection probability $|c_l|^2$ on the $\{|l,0\rangle\}$ basis for the ground state as a function of $q/|c_2|$.
 The inset shows the gap between the first excited state and the ground state. ($N=10000$).
 ({\it D}) Evolution of the normalized populations $\rho_{m_F}$ during the nonlinear $q$-ramp. Markers denote the experimental data, with solid squares, open and filled circles respectively for the $m_F = 0,1,-1$ spin components.  The open and filled circles overlay almost completely, as a result of their correlated generation. The solid lines (grey shaded regions) denote the theoretical averages (uncertainties) without fitting parameters. The insets show the ramping profile of $q$ and three typical absorption images of the atoms after Stern-Gerlach separations. The vertical dashed-line denotes the position of the QCP.
} \label{Preparation}
\end{figure}

To prepare the balanced spin-1 Dicke state $|N,0\rangle$, we resort to adiabatic approach \cite{zhangduan2013PRL}
by introducing an extra term $-q \hat N_0$ to the original Hamiltonian using electromagnetic fields, where
$q = (\epsilon_{+1}+\epsilon_{-1})/2-\epsilon_0$ denotes the effective quadratic Zeeman shift (Fig.\,1{\it B}),
with $\epsilon_{m_F}$ and $\hat N_{m_F}$ being the energy and the atom number operator for spin component $m_F$.
The linear Zeeman shift is irrelevant because the system magnetization $ \propto {\hat L_z} = {\hat N_{ + 1}} - {\hat N_{ - 1}}$
remains conserved.
When $q \gg |c_2|$, the quadratic Zeeman shift dominates, and the ground state is the polar state
with all atoms condensed in the $m_F = 0$ component.
If $q$ is adiabatically ramped to zero, an initial polar state condensate will stay in the instantaneous ground state
(within the $m=0$ subspace) and evolve into the balanced spin-1
Dicke state $|N, 0\rangle$ \cite{zhangduan2013PRL}.

The competition between spin-dependent interaction $|c_2|$ and quadratic Zeeman shift $q$ results in changing ground-state phases and a quantum critical point (QCP) at $q/|c_2|=2$ \cite{zhangduan2013PRL}.
This is clearly visible from the distinctive projected distributions $\{|c_l|^2\}$ of the ground state
$|{\rm GS}\rangle= \sum\nolimits_l {{c_l}|l,0\rangle }$ onto the zero magnetization Dicke state basis $\{|l,0\rangle\}$ on the two sides of the QCP (Fig. 1{\it C}).
When $q/|c_2|>2$, the distribution of $|c_l|^2$ is concentrated around $l \simeq \sqrt{2N}$, which gives $l/N \simeq 0$ for large $N$.
On the other side, $0\le q/|c_2|<2$, the distribution peaks approximately at $l \simeq N\sqrt{1-q^2/({2|c_2|})^2}$, which gives $l=N$ at $q=0$.

Our experiment typically starts with a condensate of
$12300\pm200$ atoms in the $m_F = 0$ component with no discernable thermal fraction
at a magnetic field of $B_0=0.815(1)\,$G (corresponding to $q=17.3|c_2|$) ({\it Methods}).
The value of $q$ is first linearly ramped to $2.7|c_2|$ in $300\,$ms, and then to zero in $1.5\,$s
by controlling the power of a dressing microwave \cite{Luo2017TwinFock}.
The energy gap between the ground and the first excited state of the system
near the QCP is less than a hertz in our case (inset of Fig.\,1{\it C}), excitation is therefore unavoidable over the finite ramp
time given the limited condensate lifetime of $\sim 30$\,s.
Optimizing the sweeping procedure thus constitutes a crucial step for the experimental success.
The ramping profile adopted (inset of Fig. 1{\it D}) is optimized first by numerical simulations,
and then fine-tuned experimentally ({\it Methods}).
At the end of the ramp, the condensate is released from the optical trap
and subjected to a pulsed gradient magnetic field, after which spin-resolved atomic populations $N_{m_F}$ are obtained with
precise absorption imaging.

The evolution of the normalized populations,
$\rho_{m_F}= N_{m_F}/N$, during the $q$-ramp is shown in Fig. 1{\it D}.
The experimental results, plotted as markers with error bars, are found to be in
excellent agreement with theoretical expectations, in solid lines for the mean values and grey shaded regions for the standard deviations,
based on solving the Hamiltonian with the experimentally-adopted ramping profile \cite{Luo2017TwinFock}.
In the first 425\,ms of the ramp, before $q$ reaches the QCP, the quadratic Zeeman energy prevails
and very few atoms are observed in $m_F = \pm 1$ (black open and red filled circles).
After crossing the QCP, the spin-mixing interaction takes over and
atoms in $m_F = \pm 1$ proliferate at the expense of those in $m_F = 0$ (blue squares).
In the end, nearly half of the atoms are transferred from $m_F = 0$, populating $m_F = \pm 1$ equally.
The population distribution (with 1/4, 1/2, 1/4 in $m_F=-1,0,+1$, respectively) gives a first indication that the prepared quantum states lie in the vicinity of the balanced Dicke state $|N,0\rangle$ ({\it Methods}).

\section*{Beating the SQL using spin-1 Dicke state}

\begin{figure*}[t!]
\centering
\includegraphics[width=1.4\columnwidth]{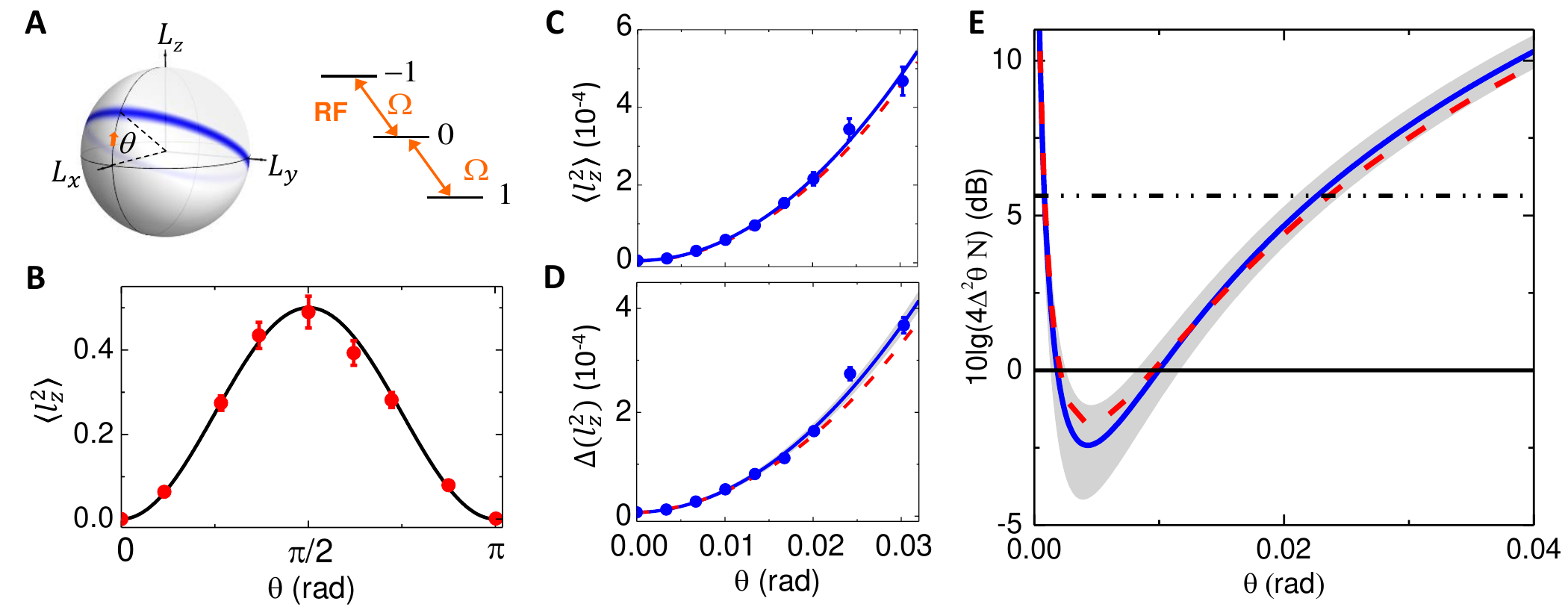}
\caption{Beating SQL with spin-1 Dicke state.
({\it A}) The balanced spin-1 Dicke state rotated by an angle $\theta$ as illustrated on the generalized Bloch sphere. The rotation is facilitated by RF Rabi coupling of equally spaced ($q=0$) spin-1 substates.
({\it B}) The second moment $\langle l_z^2 \rangle$ (red filled circles) for the rotated state as a function of the rotation angle $\theta$. The theoretical expectation is in black solid line.
({\it C}) and ({\it D}) The measured $\langle l_z^2 \rangle$ and $\Delta (l_z^2)$ as a function of $\theta$, respectively. The solid blue lines are polynomial fits.
({\it E}) Enhancement of the measured angular sensitivity using spin-1 Dicke state over the three-mode SQL.
The blue solid line is obtained from error propagation using fitting parameters from ({\it C}) and ({\it D}). The horizontal dotted-dashed line represents the two-mode SQL.
 All red dashed lines denote theoretical expectations based on measured $(\Delta {\hat{L}}_{z})_{\theta=0}$ and $L_\mathrm{eff}$ ({\it SI Appendix}). All grey shaded regions represent 1 standard deviation (s.d.) uncertainties from fitting.
 } \label{sensitivity}
\end{figure*}

The prepared spin-1 Dicke states enable a measurement precision beyond the three-mode SQL. The measurement sequence to show this is analogous to that applied in \cite{Klempt2011ScienceTF} to a spin-1/2 twin-Fock state, but involves all three $|F=1,m_F=0,\pm1\rangle$ components. We perform a well calibrated Rabi rotation of the state (equivalent to the accumulative effect of a three-mode Ramsey interferometer) by coupling the three $m_F$ states using a radio-frequency (RF) field, keeping $q=0$ (Fig.\,2{\it A}).
The rotation angle $\theta$ is then estimated from
the second moment of the measured ${\hat{l}_z}\equiv\hat{L}_z/N=(\hat{N}_{+1}-\hat{N}_{-1})/N$ (Fig.\,2{\it B}, red filled circles),
which depends on $\theta$ as $\langle {\hat l_z^2}\rangle_\theta = \sin^2(\theta)/2$ (black solid line).
From error propagation, the corresponding measurement uncertainty is given by $\Delta \theta  = \Delta({\hat l_z}^2)_\theta/|d\langle {\hat l_z}^2\rangle_\theta /d\theta |$,
which depends on the slope of the second moment $|d\langle {\hat l_z}^2\rangle_\theta /d\theta |$ and its s.d. $\Delta({\hat l_z}^2)_\theta=(\langle {\hat l_z}^4\rangle_\theta-\langle {\hat l_z}^2\rangle_\theta^2)^{1/2}$.
For small rotation angles, the measured $\langle {\hat l_z}^2 \rangle_\theta$ (Fig.\,2{\it C}) and $\Delta({\hat l_z}^2)_\theta$ (Fig.\,2{\it D}) are fitted with polynomials of $\sin(\theta)$ ({\it SI Appendix}). Using the fitting results,
an estimation of the interferometric sensitivity is obtained (Fig.\,2{\it E}).
The optimal measured sensitivity is found to lie at $\theta=0.0043\,$rad, with a value of
$\xi^2=-20 \log_{10}(\frac{\Delta \theta}{1/(2\sqrt{N})})\simeq 2.42^{+1.76}_{-1.29}\,$dB below the three-mode SQL of $1/(2\sqrt{N})$ ($N \sim 11700$ at the end of the ramp),
or $8.44^{+1.76}_{-1.29}\,$dB below the two-mode SQL of $1/\sqrt{N}$. The three-mode SQL
can be reached by using the polar state $|N_{+1}=0,N_{0}=N,N_{-1}=0\rangle$ with all $N$ atoms in $m_F=0$ as input \cite{kajtoch2018metrologically} ({\it Methods}).

The experimentally achievable best interferometric sensitivity of the prepared states is limited by $(\Delta \theta)_\mathrm{opt} = [{ 3(\Delta \hat L_{z})^2_{\theta=0} +1/2 }]^{\frac{1}{2}}/L_{\rm eff}$ ({\it SI Appendix}). Here, $L_{\rm eff} = \langle { {\hat {\bf L}^2} } \rangle^{\frac{1}{2}}$ and ${(\Delta \hat L_z)}_{\theta=0}$ are determined from measurements.
With an ideal balanced spin-1 Dicke state and perfect detection, we would have ${(\Delta \hat L_z)}_{\theta=0}=0$ and $L_{\rm eff}= \sqrt{N(N+1)}$. However, due to atom loss and  detection noise, we measure a ${(\Delta \hat L_z)}_{\theta=0} = 25.51\pm 0.55$ from 1000 consecutive samples (left panel of Fig.\,3{\it A}). Although nonzero, this measured value is much smaller than the transverse spin uncertainty or the quantum shot noise (QSN) of $\sqrt{N} \simeq 108.17 \pm 0.72$ for the polar state,
and gives a number squeezing of $\xi_N^2 = -20 \log_{10} [\frac{{(\Delta \hat L_z)}_{\theta=0}}{\sqrt{N} }]\simeq 12.56\pm0.45\,$dB below the QSN. After subtracting the quantitatively well understood detection noise ($\Delta {\hat L_z}^{\rm DN} = 21.4$), we infer a number squeezing of $17.83\pm 1.48\,$dB.

The effective spin length $L_{\rm eff}$ for the prepared Dicke states is
determined from $\hat L_z$ measurement after they are rotated to the vertical direction ($\theta=\pi/2$ as shown in Fig.\,3{\it A})({\it Methods}).
The right panel of Fig.\,3{\it A} shows the histogram of the measured $L_z/N$
for 1043 continuous runs. The measured distribution matches well to the theoretical expectation (black solid line)
of the target Dicke state $|N,0\rangle$. It infers a normalized collective spin length squared of $\langle \hat {\bf L}^2 /[N(N+1)]\rangle=1.000\pm0.021$ ({\it Methods}).
With such a nearly perfect coherence, the optimal achievable phase squeezing $\xi^2_\mathrm{opt}=-20\log_{10} (\frac{\Delta \theta_\mathrm{opt}}{1/(2\sqrt{N})})$ is limited by the number squeezing $\xi^2_N$, and is given by
$\xi^2_\mathrm{opt}\simeq \xi_N^2 - 10\log_{10}12 \simeq 1.77\,$dB ({\it SI Appendix}).
This value agrees with the observed phase squeezing of $\xi^2=2.42^{+1.76}_{-1.29}$dB.
Combining the measured normalized collective spin length and the detection-noise-subtracted number squeezing of $17.83\pm 1.48\,$dB,
 we can infer an entanglement breadth of more than $10000$ 
 atoms on average, and at least $\approx 630$ atoms at 1 s.d. according to the criterion of refs. \cite{Klempt2014PhysRevLett.112.155304,Molmer2001PhysRevLett.86.4431,vitagliano2017entanglement} ( Fig.\,3{\it B}).

We now contrast our results with three related works based also on spin-mixing dynamics of $^{87}$Rb BEC \cite{Klempt2011ScienceTF,Klempt2016PRL117.143004, Oberthaler2016PRL.117.013001}. In \cite{Klempt2016PRL117.143004}, a squeezed vacuum state with mean occupation of 0.75 atoms is prepared in the $m_F=\pm1$ components. Using this state (with $\approx10000$ atoms in the $m_F=0$ component) as input for a two-mode Ramsey interferometer, a phase measurement precision $2.05^{+0.34}_{-0.37}$\,dB beyond the two-mode SQL of 10000 atoms is demonstrated. In \cite{Oberthaler2016PRL.117.013001}, Linnemann et al. realize a SU(1,1) interferometer by using the spin-mixing dynamics to act as nonlinear beam-splitter.  With this, they demonstrate interferometric sensitivity beyond the two-mode SQL of $\approx$2.8 atoms. L{\"u}cke et al. \cite{Klempt2011ScienceTF} beat the two-mode SQL by $1.6^{+0.98}_{-1.1}$\,dB using post-selected twin-Fock states of about 7000 atoms. In comparison, our deterministically prepared spin-1 Dicke states beat the three-mode SQL of 11700 atoms by $2.42^{+1.76}_{-1.29}\,$dB and two-mode SQL by $8.44^{+1.76}_{-1.29}\,$dB. The measured interferometric sensitivities of both \cite{Klempt2011ScienceTF} and our work are mainly limited by atom number detection resolutions, instead of the qualities of the prepared states.

\section*{Benchmarking the prepared spin-1 Dicke state}

\begin{figure*}[t!]
\centering
\includegraphics[width=1.6\columnwidth]{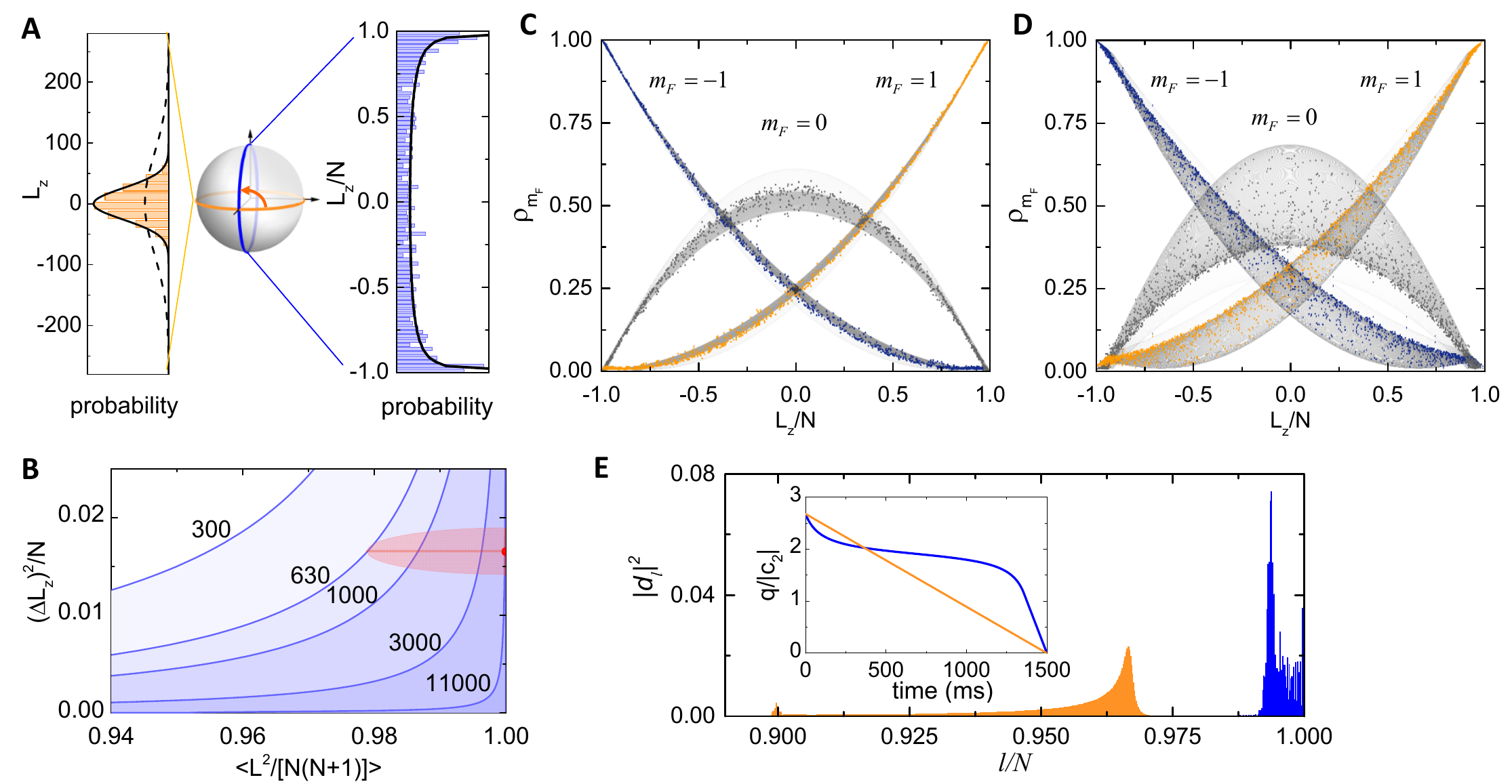}
\caption{ {\bf Benchmarking the prepared spin-1 Dicke states.}
({\it A}) Histogram of measured $L_z$ for the prepared Dicke states (left panel, in light orange) and that after a $\pi/2$ rotation (right panel, in light blue), with their corresponding representations on the Bloch sphere shown in the middle panel. The black dashed line in the left panel denotes the distribution expected for the ($\pi/2$ rotated) polar state. The black solid line in the right panel denotes the theoretical expectation for the ideal Dicke state $|N,0\rangle$.
({\it B}) Analysis of entanglement breadth for the spin-1 Dicke samples using the measured normalized collective spin length and detection-noise-subtracted number squeezing, following refs. \cite{Klempt2014PhysRevLett.112.155304,Molmer2001PhysRevLett.86.4431,vitagliano2017entanglement}. A state below the boundary labeled with number $k$ contains at least a group of non-separable $k$ particles. The solid red circle (only the left half is visible) denotes the result for our samples.  The red ellipse represents uncertainties of the measurements at 1 s.d..
({\it C}) The distribution of the prepared states after a $\pi/2$ rotation in the Fock state basis.
The solid dots denote the measured data of occurrences, obtained
from 1043 continuous experiment runs, while the grey shaded regions represent theoretical occurrence probability.
({\it D}) The same as in ({\it C}) but for a linear ramp and from 2000 continuous experiment runs.
({\it E}) The theoretical distribution of the prepared states $|\psi\rangle$ on the Dicke basis $\{|l,0\rangle\}$, $|\psi \rangle  = \sum\nolimits_l {{d_l}|l,0\rangle } $, are compared using linear and nonlinear $q$-ramp, with the profiles of the ramps shown in the inset.
 } \label{IdentifyDicke}
\end{figure*}

We now characterize the quality of the state we prepare in comparison to the ideal balanced Dicke state $|N,0\rangle$.
Fig. 3{\it C} shows the distribution of the prepared states after ($\theta=\pi/2$) rotation in the Fock state basis (using the same data for the right panel of Fig.\,3{\it A}). Its main features can be understood by considering the ideal state $|N,0\rangle$. The $\pi/2$ rotation turns $|N,0\rangle$ into a superposition of Dicke states of different $m$, $|N,m\rangle$, whose projections in the Fock basis are given by $\sum_k f_k^{(N,m)} |k+m, N-2k-m, k\rangle$. Measurements which give $L_z=m$ therefore show an outcome of $\rho_{+1}=\frac{k+m}{N}$, $\rho_{0}=\frac{N-2k-m}{N}$, and $\rho_{-1}=\frac{k}{N}$ with relative probability of $|f_k^{(N,m)}|^2$. For a fixed $m$, the resulting probability distribution for $\rho_0$ is well approximated by a Gaussian function centered at
$\frac{1}{2}(1-\frac{m^2}{N^2})$ with a width of $\Delta \rho_0 = \frac{1}{2}\sqrt{(1-m^2/N^2)/N}$ ({\it SI Appendix}).
The experimental results show a similar structure but with a larger width. The discrepancy stems from populating of the excited Dicke states $|l,0\rangle$ ($l<N$) while crossing the QCP. This conclusion is supported by the good agreement between the observed distributions and the expected ones from our simulation of the prepared Dicke state $|\psi \rangle  = \sum\nolimits_l {{d_l}|l,0\rangle }$ ({\it SI Appendix}). The reliability of our analysis is further confirmed in another set of experiments using a linear (less adiabatic) ramp of $q$ (Fig.\,3{\it D}), which show even broader distributions due to expected higher excitations.

We find that the observed peak-to-peak spread of $\rho_0$ at $m=0$ can be used to determine the highest excitation (or minimum $l_\mathrm{min}$) by $\rho_{0(pp)}\simeq\sqrt{2(1-l_\mathrm{min}/N)}$ ({\it SI Appendix}).
For the nonlinear ramp, we infer the highest excitation with $l_\mathrm{min}/N\approx0.99$, i.e.,
the prepared state occupies only $|l,0\rangle$ states with $0.99\lesssim l/N\leq1.00$.
For the linear ramp, an excitation upper bound at $l_\mathrm{min}/N\approx0.90$ is inferred instead.
These estimates agree well with the excitation spectra $|d_l|^2$ from theoretical simulations (Fig.\,3{\it E}).

In conclusion, we report the first deterministic preparation of high quality balanced spin-1 Dicke states, by driving
a condensate of $^{87}$Rb atoms through a QCP. The prepared states are used to demonstrate a rotation measurement sensitivity of $2.42^{+1.76}_{-1.29}\,$dB ($8.44^{+1.76}_{-1.29}\,$dB)
beyond the three-mode (two-mode) SQL, limited by the atom number detection resolution.
We anticipate that our work will stimulate the experimental pursuit of higher-spin entangled states besides the widely explored spin-1/2 ones.

This work is supported by NSFC (No. 91421305, No. 91636213, No. 11654001, No. 91736311, and No. 11574177), and by the National Basic Research Program of China (973 program) (No. 2014CB921403).
\\

\noindent Author contributions: YQZ, QL, XYL, SFG, JHC carried out experimental work and data analysis. The theoretical work was performed by LNW. LY and MKT supervised the work. YQZ, LNW, MKT and LY wrote the manuscript. YQZ and LNW contributed equally to this work.
%

\section{Methods}

\subsection{ Main experimental sequence}
A condensate of about $1.2\times10^4$ $^{87}$Rb atoms in the $5$s $\left| {F=1,m_F=0} \right\rangle$ hyperfine ground state is prepared
inside an optical dipole trap formed by two crossed $1064$-nm light beams following procedures described in ref. \cite{Luo2017TwinFock}.
The atoms are under a bias magnetic field, along the gravity direction and fixed at $0.815$\,G, actively stabilized to a r.m.s uncertainty of $20\,\mu$G with a fluxgate magnetometer. The optical trap is then compressed to the final trapping frequencies of $2\pi \times (210,108,169)$\,Hz in $300$\,ms giving a spin mixing rate $c_2=-2\pi \times 2.75(2)$\,Hz. Over the same $300$-ms, $q$ is ramped from $17.3|c_2|$ to $2.7|c_2|$ with a dressing microwave. The main experiment starts by ramping $q$ from $2.7|c_2|$ to $0$ in $1.5$\,s with an optimized profile to generate spin-1 Dicke state. Rotation of the spin-1 Dicke state is performed using a radio-frequency pulse resonant to both the $|F=1,m_F=0\rangle$ to $|F=1,m_F=\pm1\rangle$ splittings, always keeping $q=0$. At the end of the experiment, the trap is switched off abruptly and atoms in different $m_F$ states are separated by the Stern-Gerlach technique over a time of flight of $8$\,ms, after which, absorption images record atoms in all three spin components. Details about the low noise detection and calibration of atom numbers are as outlined in ref. \cite{Luo2017TwinFock}.
\subsection{Calibrating $q$ and $c_2$}
The effective quadratic Zeeman shift  $q = q_B+q_M$ is determined by the quadratic Zeeman shift $q_B$ and the microwave induced ac-Zeeman shift $q_M$. In our experiment, the static magnetic field is fixed. Tuning of $q$ is accomplished with a dressing microwave which is $19$\,MHz blue-detuned to the $\left| {F=1,m_F=0} \right\rangle$ to $\left| {F=2,m_F=0} \right\rangle$ transition of the $^{87}$Rb atoms. Here $q$ varies linearly with the microwave power according to a setup-specified slope that is precisely calibrated as in ref. \cite{Luo2017TwinFock}. Another important parameter $c_2=-2\pi \times 2.75(2)$\,Hz is also precisely calibrated using the method described in ref. \cite{Luo2017TwinFock}. Our experiment demands extreme stability for these two parameters. The microwave power is controlled to a stability at the level of one thousandth. All data collections are carried out after two-hour warm-up of the
 experimental setup when the drifts for $c_2$ and $q$ become less than 1\%.
\subsection{Ramping profile}
The ramping profile of $q$ we use is designed with the main aim of minimizing the excitation of the system and atom loss. The basic idea is to ramp slower across the QCP, where the energy gap is the smallest and where excitations occur most easily. For our case,
the ramping profile is optimized based on the following form,
\begin{equation}\label{qt}
\frac{q\left( t \right)}{|c_2|} = \varepsilon  \cdot \left\{ {\begin{array}{*{20}{l}}
{ - \beta \tan \left[ {2\left( {\frac{t}{\tau } - \frac{1}{2}} \right){{\tan }^{ - 1}}\left( {\frac{{{q_m} - {q_c}}}{\beta }} \right)} \right]{\rm{ + }}{q_c},\quad 0 \le t \le \tau .}\\
{k\left( {t - \tau } \right) + 2{q_c} - {q_m},\quad \tau  < t \le {t_f}.}
\end{array}} \right.
\end{equation}
It consists of two parts: the first part is a modified tangent function \cite{bason2012high} which features a gentler slope near the QCP ($q_c = 2$); the second part is a linear function whose slope $k$ is fixed by the conditions that the two functions and their derivatives are continuous and is given by
\begin{equation}
k = \frac{{\beta \left( {{q_m} - 2{q_c}} \right) + 2\left[ {{\beta ^2} + {{\left( {{q_c} - {q_m}} \right)}^2}} \right]{{\tan }^{ - 1}}\left[ {\left( {{q_c} - {q_m}} \right)/\beta } \right]}}{{\beta {t_f}}}.
\end{equation}
The total ramp time $t_f$ of $1.5$\,s represents a compromise between adiabaticity and atom loss. The latter is detrimental to the achievable measurement precision. The duration of the first part of the ramp $\tau$ is determined by the condition that $\tau-(2q_c-q_m)/k=t_f$, i.e., the sum of the two durations gives the total evolution time $t_f$, and is given by
\begin{equation}
\tau  = \frac{{{t_f}}}{{1 + \frac{{\beta \left( {{q_m} - 2{q_c}} \right)}}{{2\left[ {{\beta ^2} + {{\left( {{q_c} - {q_m}} \right)}^2}} \right]{{\tan }^{ - 1}}\left[ {\left( {{q_c} - {q_m}} \right)/\beta } \right]}}}}.
\end{equation}
The remaining two parameters in Eq. (\ref{qt}), i.e., the steepness of the tangent function $\beta$ and the start point of the ramp $q_m$, are optimized by numerical simulation to give the largest effective collective spin length $L_{\rm eff}$ (or equivalently the least excitations).
Taking into account several factors such as atom loss in the experiment, an overall scaling parameter $\varepsilon$
is introduced in Eq.~(1), which is optimized experimentally.
Eventually, the parameters we use are $\beta = 0.16$, $q_m = 2.8$ and $\varepsilon = 0.956$.
\subsection{Measurement of $L_\mathrm{eff}^2$}
The squared effective spin length is $L_{\rm eff}^2 = \langle { {\hat {\bf L}^2} } \rangle =\langle \hat{L}_x^2+\hat{L}_y^2+\hat{L}_z^2 \rangle$.
To measure $L_x$ or $L_y$, we rotate the state by $90^{\circ}$ using a $\pi/2$ RF pulse before the $L_z$ measurement \cite{Klempt2014PhysRevLett.112.155304}.
The squared effective spin length is then calculated from the measured $L_z^2$
as $L_{\rm eff}^2 =2\langle \hat{L}_z^2 \rangle$.
The statistical uncertainties are calculated based on the estimators given in ref. \cite{Klempt2014PhysRevLett.112.155304}.
\subsection{Balanced spin-1 Dicke state and coherent spin states}
The balanced spin-1 Dicke state can be represented as a superposition of coherent spin states equally distributed on the equator of the Bloch sphere,
$\propto \int_0^{2\pi } {d\varphi {{\left( {\frac{1}{2}{e^{ - i\varphi }}|{m_F} = 1\rangle  + \frac{1}{{\sqrt 2 }}|{m_F} = 0\rangle  + \frac{1}{2}{e^{i\varphi }}|{m_F} =  - 1\rangle } \right)}^{ \otimes N}}} $. The above relation represents a spin-1 extension of a spin-1/2 fragmented BEC \cite{Mueller2006PhysRevA.74.033612}. It is worth emphasizing that, in most observations, a balanced Dicke state appears to be very similar to coherent spin states lying on the equator of the Bloch sphere albeit with random phases. However, the quantum noise of these coherent states cancels each other out along the $\hat{L}_z$ direction, resulting in zero fluctuations in $\hat{L}_z$ measurements.
\subsection{Squeezing limitations}
The number squeezing is mainly limited by detection noise and atom loss. The largest contribution is the atom-independent detection noise which arises mostly from the photon shot noise of the probe light and amounts to $\Delta \hat L_z^{\rm DN} = 21.4$. Another sizable factor is atom loss in the $m_F=\pm1$ components. During the ramp, around $\eta =5\%$ of the total atoms are lost. It sets a noise floor of $\Delta \hat L_z^{\rm loss} \simeq\sqrt{N\eta\alpha}\simeq 5.7$, where $\alpha  = \frac{1}{t_f}\int_0^{t_f} {{(\langle \rho_1 \rangle + \langle \rho_{-1} \rangle)}dt} \simeq 0.06$ is the average ratio for the atoms being in the spin components $m_F=\pm1$.
Combining these two factors, we expect a number squeezing of $13.7$\,dB. We attribute the slight difference between this and our measured value of $12.56$\,dB to atom-number-dependent technical noise, whose origin deserves further investigation.
\subsection{SQL of a $M$-mode interferometer}
The optimal phase sensitivity for an interferometer is given by
$\Delta \theta = 1/\sqrt{F_Q}$, where $F_Q$ denotes the
quantum Fisher information (QFI) and depends on the interferometric operation and
the input state. When a pure state is fed into a linear interferometer operationally described by $\exp{(-i \hat h \theta)}$, the QFI is given by $F_Q = 4 (\Delta \hat h)^2 = 4(\langle\hat h^2\rangle-\langle\hat h\rangle^2)$.
Its upper
bound (optimized over all states obtained from the input state through linear operations)
is provided by [31] $F_Q =(\lambda_{\rm max} - \lambda_{\rm min})^2$,
where $\lambda_{\rm max}$ and $\lambda_{\rm min}$ respectively correspond to the largest and the smallest eigenvalues of $\hat h$.
For the rotation operation we consider, the phase generator $\hat h$ corresponds to the
collective spin component $\hat L_y$. For a single-particle (with spin $s$) state fed into a $M$($=2s+1$)-mode interferometer, the optimal QFI becomes $(s-(-s))^2=(2s)^2=(M-1)^2$. Hence, the SQL for $N$ such particles is given by $1/[(M-1)\sqrt{N}]$. A more general proof for the M-mode SQL can be found in ref.\,\cite{Smerzi2013PhysRevA.87.033607}.

\subsection{Phase sensitivities for Dicke state and Polar state}
The QFI for the Dicke state $|l,m\rangle$ is given by $F_Q = (\Delta \hat L_y)^2 = 2[l(l+1)-m^2]$.
Therefore, the highest precision $1/\sqrt{2N(N+1)}$ comes from the
balanced Dicke state $|l=N,m=0\rangle$.
The QFI for the polar state $|0,N,0\rangle$ is given by $4(\Delta \hat L_y)^2 = 4N$. This leads to an optimal precision of $1/(2\sqrt{N})$, saturating the 3-mode SQL.

In this report, $A$ without hat represents measured values of the operator $\hat{A}$.

\clearpage
\section{Supporting Information}

\subsection{Small-angle rotation}
In the presence of detection noise, the best phase sensitivity is achieved at a small rotation angle about a few milli radians in our case. Such a small rotation angle $\theta =\Omega \tau$ implicates small Rabi frequency $\Omega\propto \sqrt{P}$ (power $P$ of the RF field) or short pulse duration $\tau$.
However, RF pulses with small Rabi frequencies are sensitive to magnetic field noise (the bias magnetic field suffers a shot-to-shot noise of $20\,\mu$G (r.m.s.) in our experiment), and short pulses are subjected to difficult-to-control switching effects. To overcome these problems, instead of directly performing a small-angle rotation, we use more precisely controlled composite pulses, which feature a forward rotation followed by a backward rotation. For instance, a 10-$\mu$s forward rotation of $0.063$\,rad followed by a 10-$\mu$s backward rotation of $-0.060$ rad gives a net rotation of $0.003$\,rad.

\begin{figure}
\centering
  \includegraphics[width=0.75\columnwidth]{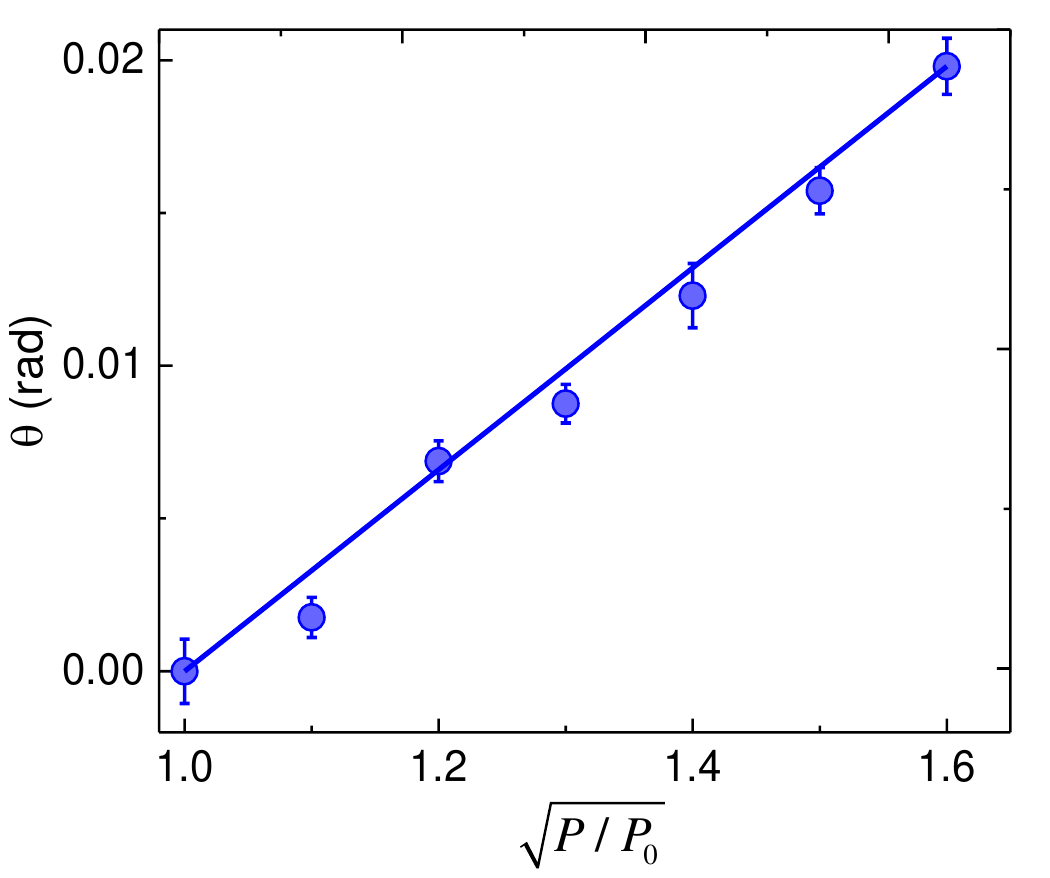}
  \caption{Calibration of the rotation angle. The dependence of rotation angle $\theta$ on the relative strength of the second pulse. The blue solid line is a linear fit of the experimental results averaged over $20$ runs for each point.
  }\label{Rotation_Cali}
\end{figure}

In our experiment, two 10-$\mu$s RF pulses with $\Omega \sim 700\,$Hz, whose relative phase is set at $\pi$, are used. By slightly changing the power of the second pulse, the net rotation angle can be controlled. As the pulse contains about only 5 periods of oscillations, we reset the phase accumulator of the direct digital synthesizer (DDS) to make sure that the RF waveform stays the same for repeated experiments.
As the duration of the composite pulse is short, the switching effect cannot be neglected.
Hence, the rotation angle of each composite pulse has to be calibrated independently by
measuring the population imbalance ${\hat L_z}={\hat N_{+1}}-{\hat N_{-1}}$ of the composite-pulse-rotated coherent spin state
$e^{i\theta \hat L_y}(\frac{1}{2}|1,-1\rangle+\frac{1}{\sqrt{2}}|1,0\rangle+\frac{1}{2}|1,+1\rangle)^{\otimes N}$, which depends on the rotation angle $\theta$ as
$\langle \hat L_z \rangle = N \sin\theta \simeq N \theta$ for small $\theta$. The calibration procedure starts with the preparation of all atoms (around 60000) in spin component $\left| {F=1,m_F=-1} \right\rangle$, which are then rotated by the composite pulse, followed by a $\frac{\pi}{2}$-pulse-rotation. The population imbalance $L_z$ is then recorded. The averaged value from repeated experiments determines the rotation angle of the composite pulse $\theta$\,($=\langle \hat L_z /N \rangle$). This procedure is repeated
with the power of the first pulse in the composite pulse $P_0$ fixed while the second one $P$ varied.
The dependence of the rotation angle $\theta$ on the relative strength of the second pulse defined as $\sqrt{(P/P_0)}$ is shown in Fig.~\ref{Rotation_Cali}. The fitting curve gives $\theta = 0.033\times[\sqrt{(P/P_0)}-1]$, which is then used to determine the relative strength of the second pulse given an angle $\theta$ in the following experiments.

\subsection{Entanglement breadth}
\begin{figure}
\centering
  \includegraphics[width=1\columnwidth]{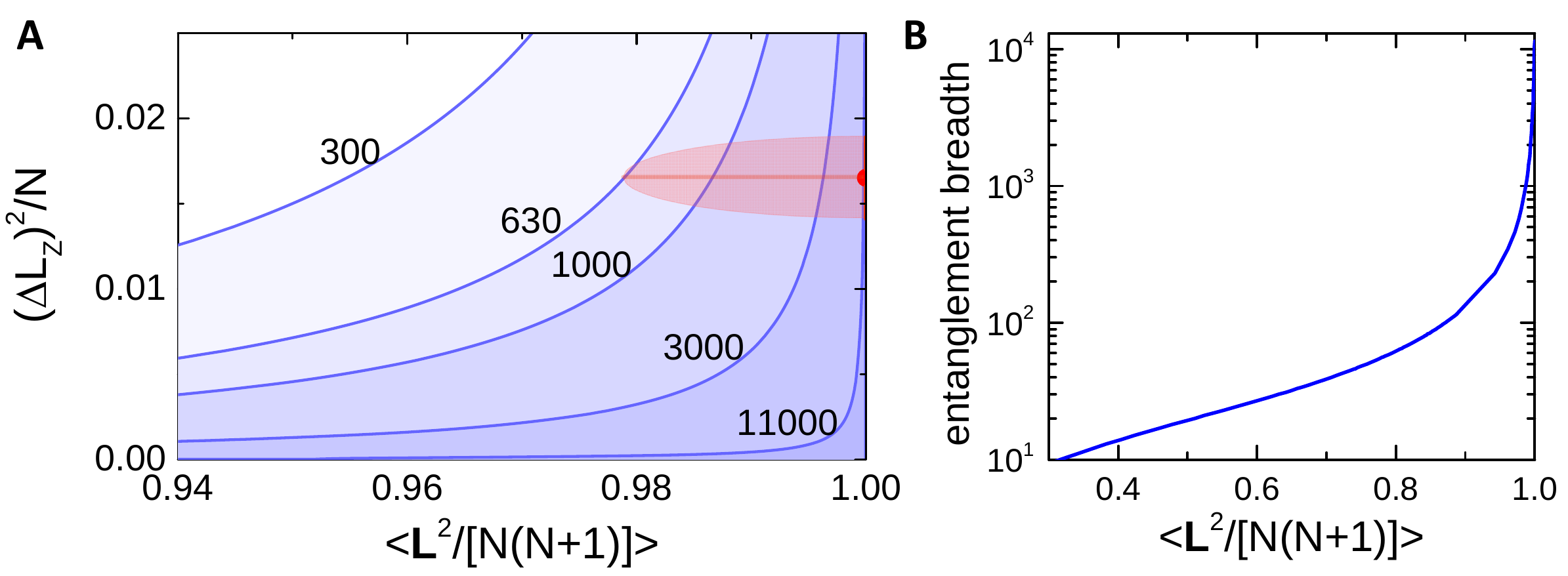}
  \caption{ Entanglement breadth. ({\it A}) Analysis of entanglement breadth for the prepared samples following ref~\cite{Klempt2014PhysRevLett.112.155304}. A state below
    the boundary (blue line) labeled with number $k$ contains at least a subgroup of nonseparable $k$ particles. The tiny solid red circle
    (only the left half is visible) denotes the result for our samples, which gives $(\Delta \hat L_z)^2/N = 0.0165 \pm 0.0024$ and $\langle \hat {\bf L}^2/[N(N+1)] \rangle = 1.000\pm0.021$. The red ellipse represents uncertainties of the measurements at $68.3\%$ statistical confidence interval. ({\it B}) The cut along the horizontal red solid line in ({\it A}) shows the extremely steep dependence of the entanglement breadth on the effective spin length given $(\Delta \hat L_z)^2/N = 0.0165$.}
    \label{entangle}
\end{figure}

For states in the vicinity of Dicke states, ref.~\cite{Klempt2014PhysRevLett.112.155304} proposes a method to determine
their entanglement breadth based on the measurement of the collective spins. Although the original
criterion is for spin-$1/2$ states, it can be generalized to spin-1 ones by replacing the
collective spin-$1/2$ operators by the corresponding collective spin-$1$ ones. Hence,
following the analysis of ref~\cite{Klempt2014PhysRevLett.112.155304}, we can infer the entanglement breadth of our samples
from the detection-noise-subtracted normalized $\hat L_z$, $(\Delta \hat L_z)^2/N$, and the normalized
squared effective spin length, $\langle \hat {\bf L}^2/[N(N+1)] \rangle$, as shown in Fig.\,\ref{entangle}{\it A}.
The boundary labeled by number $k$ is given by the state
\begin{equation}\label{state}
|\Psi \rangle  = |\psi_k {\rangle ^{ \otimes n}} \otimes |\psi_p \rangle,
\end{equation}
which is a product of $n$\,($=\left\lfloor {N/k} \right\rfloor $, integer part of $N/k$) copies of state $|\psi_k \rangle $ containing $k$ nonseparable spin-$1$ particles and state $|\psi_p \rangle $ composed of the remaining $p$\,($=N-nk$) particles.
The state $|\psi_{\mu}\rangle$\,($\mu=k,p$) represents the ground state of the Hamiltonian
\begin{equation}\label{h}
{H_\mu } = \hat L_z^{(\mu )2} - \lambda \hat L_x^{(\mu )},
\end{equation}
where ${\hat {\bf L}^{(\mu )}} = \sum\nolimits_{j = 1}^\mu  {{\hat {\bf F}^{(j)}}} $ is the collective spin-$1$ operator. The boundary points (solid lines in Fig.\,\ref{entangle}{\it A}) are obtained as
\begin{eqnarray}\label{b}
\langle {\hat L}_x^2 + {\hat L}_y^2\rangle  &=& n{\langle {\hat L}_x^{(k)2} + {\hat L}_y^{(k)2}\rangle _{|{\psi _k}\rangle }} + n\left( {n - 1} \right)\langle {\hat L}_x^{(k)}\rangle _{|{\psi _k}\rangle }^2 \notag\\
&&+ {\langle {\hat L}_x^{(p)2} + {\hat L}_y^{(p)2}\rangle _{|{\psi _p}\rangle }} + 2n{\langle {\hat L}_x^{(k)}\rangle _{|{\psi _k}\rangle }}{\langle {\hat L}_x^{(p)}\rangle _{|{\psi _p}\rangle }}, \notag\\
{( {\Delta {{\hat L}_z}} )^2} &=& n ( {\Delta {\hat L}_z^{(k)}} )_{|{\psi _k}\rangle }^2 + ( {\Delta {\hat L}_z^{(p)}} )_{|{\psi _p}\rangle }^2.
\end{eqnarray}

Fig.\,\ref{entangle}{\it B} shows the dependence of the entanglement breadth on the effective spin length, given $(\Delta \hat L_z)^2/N = 0.0165$. The slope of the dependence diverges as the effective spin length approaches unity. This explains the inferred entanglement breadth is about 10000 atoms on average and $\approx$ 630 atoms at $68.3\%$ statistical confidence interval.

\subsection{Interferometric sensitivity}
Here, we show how we obtain the results for the blue solid and red dashed lines in Figs. 2{\it C}, 2{\it D} and 2{\it E} in the main text.

In the presence of atom loss and when the ramp of $q$ is nonadiabatic, the prepared state is a
mixture of Dicke states $\rho  = \sum\nolimits_{l,m} {{p_{l,m}}|l,m\rangle \langle l,m|}$.
After a Rabi rotation of angle $\theta$, the state becomes $\rho_\theta = {\hat U_\theta} \rho {\hat U_\theta}^{\dag}$, where $\hat U_\theta = \exp{(-i \theta \hat L_y})$.
It is straightforward to verify that the expectation value of ${\hat L_z}^2$ for the output state $\rho_\theta$ is given by
\begin{eqnarray}\label{Lz2mean}
\langle {\hat L_z}^2\rangle_\theta = \langle {\hat L_z}^2\rangle {\cos ^2}\theta  + \langle {\hat L_x}^2\rangle {\sin ^2}\theta ,
\end{eqnarray}
where $\langle \hat O \rangle = {\rm{tr}}(\rho \hat O)$ denotes the expectation value of operator $\hat O$ with respect to the non-rotated input state $\rho$.
The corresponding variance reads
\begin{eqnarray}\label{varLz2}
{( {\Delta {\hat L_z}^2} )^2_\theta} &\equiv &\langle {\hat L_z}^4\rangle_\theta - {\langle {\hat L_z}^2\rangle ^2_\theta} \notag\\
 &=& {( {\Delta {\hat L_z}^2} )^2}{\cos ^4}\theta  + {( {\Delta {\hat L_x}^2} )^2}{\sin ^4}\theta  + {V_{xz}}{\sin ^2}\theta {\cos ^2}\theta,\notag\\
\end{eqnarray}
with ${V_{xz}} = \langle {( {{{\hat L_x}}{{\hat L_z}} + {{\hat L_z}}{{\hat L_x}}} )^2}\rangle  + \langle {\hat L_z}^2{\hat L_x}^2 + {\hat L_x}^2{\hat L_z}^2\rangle  - 2\langle {\hat L_z}^2\rangle \langle {\hat L_x}^2\rangle $.
For the state $\rho  = \sum\nolimits_{l,m} {{p_{l,m}}|l,m\rangle \langle l,m|}$ prepared in our experiment, the populated states $|l,m\rangle$ are concentrated heavily in the region
$l \simeq N$ and $l \gg m$. In this case, we can approximate Eq. (\ref{varLz2}) by \cite{Luo2017TwinFock}
\begin{eqnarray}\label{varLz2_app}
( {\Delta {\hat L_z}^2} )_\theta^2 &\simeq& 2{\langle {\hat L_z}^2\rangle ^2}{\cos ^4}\theta+ \frac{1}{2}{\langle {\hat L_x}^2\rangle ^2}{\sin ^4}\theta   \notag\\
&&+ [ {4\langle {\hat L_z}^2\rangle  + 1} ]\langle {\hat L_x}^2\rangle {\sin ^2}\theta {\cos ^2}\theta.
\end{eqnarray}
Further assuming that the imperfect detection introduces a Gaussian noise $\sigma_{\rm dn}$, the average and standard deviation of the detected ${\hat L_z}^2$ for a small rotation angle $\theta$ are then given by~\cite{Luo2017TwinFock}
\begin{eqnarray}
{\langle {\hat L_z}^2\rangle_\theta} &\simeq & ( {\Delta {\hat L_z}} )_{{\rm{det}}}^2 + \langle {\hat L_x}^2\rangle {\sin ^2}\theta, \notag\\
( {\Delta {\hat L_z}^2} )_\theta &\simeq & \left\{{\langle {\hat L_x}^2\rangle ^2}{\sin ^4}\theta/2 + 2{( {\Delta {\hat L_z}} )_{{\rm{det}}}^4}\right.\notag\\
 &&\left.+ [{4{( {\Delta {\hat L_z}} )_{{\rm{det}}}^2} + 1} ]\langle {\hat L_x}^2\rangle {\sin ^2}\theta  \right\}^{\frac{1}{2}},\label{dLz_t}
\end{eqnarray}
respectively, where ${( {\Delta {\hat L_z}} )_{{\rm{det}}}^2} = \langle {\hat L_z}^2\rangle  + \sigma _{{\rm{dn}}}^2$ (denoted in shorthand by $(\Delta \hat L_{z})_{\theta=0}^2$ in the main text) is the fluctuation of the detected ${\hat L_z}$ for the non-rotated samples.

To determine the interferometric sensitivity, we measured $\langle \hat{l}_z^2 \rangle_\theta=\langle (\hat{L}_z/N)^2 \rangle_\theta$ and $\Delta (\hat{l}_z^2)_\theta$ for a range of small rotation angles. The normalized $\hat{l}_z$ is used instead because this gives more accurate results when the atom number $N$ varies from one experimental run to the other. The calculations of $\langle \hat{l}_z^2 \rangle_\theta$ and $\Delta (\hat{l}_z^2)_\theta$ follow the method in ref. \cite{Klempt2014PhysRevLett.112.155304} where an unbiased estimator is adopted for the second moment and the corresponding variance.
These values at all measured angles are fitted by
\begin{eqnarray}\label{fit}
{f_{\langle {\hat l_z}^2 \rangle}}\left( \theta  \right) &=& {a_1}{\sin ^2}\theta  + {b_1}, \notag\\
{f_{\Delta {\hat l_z}^2}}\left( \theta  \right) &=& \sqrt{{a_2}{\sin ^4}\theta  + {b_2}{\sin ^2}\theta  + {c_2}},
\end{eqnarray}
with
$a_1 = 0.527 \pm 0.010$, $b_1 = (5.50 \pm 0.347)\times 10^{-6}$,
$a_2 = 0.156 \pm 0.013$, $b_2 = (8.09 \pm 2.89)\times 10^{-6}$, and
$c_2 = (5.28 \pm 1.95)\times 10^{-11}$.
The average fitting functions for  $\langle {\hat l_z}^2 \rangle_\theta$ and $\Delta ({\hat l_z}^2)_\theta$ are shown as blue solid lines in Figs. 2{\it C} and 2{\it D} in the main text. The shaded regions in these figures denote the fitting uncertainties.

By substituting the measured
 $\langle {\hat L_x}^2\rangle  = {\langle {\hat {\bf L}^2}\rangle }/2 = \langle N\rangle ( {\langle N\rangle  + 1} )/2$ with $\langle N\rangle  = 11500$ (for this set of experiments)
 and ${( {\Delta {\hat L_z}} )_{{\rm{det}}}^2} = 25.5^2$ into Eq.\,(\ref{dLz_t}), the theoretical expectations of $\langle {\hat L_z}^2\rangle_\theta/\langle N\rangle^2$ and $({\Delta {\hat L_z}^2} )_\theta/\langle N\rangle^2$, based on measured properties of the non-rotated states, are shown as red dashed lines in Figs. 2{\it C} and 2{\it D} in the main text, respectively.


According to the error propagation formula, the corresponding phase sensitivities are given by
\begin{eqnarray}\label{precision}
{ {\Delta \theta } } = \frac{{ {\Delta ({\hat l_z}^2} )_\theta}}{{|d{{\langle {\hat l_z}^2\rangle }_\theta}/d\theta |}},
\end{eqnarray}
for the measurement results (blue solid line in Fig. 2{\it E}, based on Eq. \ref{fit}), and by
\begin{eqnarray}\label{precision}
{ {\Delta \theta } } = \frac{{ {\Delta ({\hat L_z}^2} )_\theta}}{{|d{{\langle {\hat L_z}^2\rangle }_\theta}/d\theta |}},
\end{eqnarray}
for the theoretical prediction (red dashed line in Fig. 2{\it E}, based on Eq. \ref{dLz_t}).

The expected optimal phase sensitivity based on the measured ${\langle {\hat {\bf L}^2}\rangle }$ and ${( {\Delta {\hat L_z}} )_{{\rm{det}}}^2}$ can be well approximated by~\cite{Luo2017TwinFock}
\begin{eqnarray}\label{sensitivity}
( {\Delta \theta } )_{{\rm{opt}}} \simeq \sqrt{\frac{{3( {\Delta {{\hat L_z}}} )_{{\rm{det}}}^2 + 1/2}}{{\langle {\hat {\bf L}^2}\rangle }}}.
\end{eqnarray}

For this experiment, the best measured interferometric uncertainty lies $2.42^{+1.76}_{-1.29}$ dB below the three-mode standard quantum limit (SQL) and $8.44^{+1.76}_{-1.29}$ dB below the two-mode SQL.
As in all experiments on measurement beyond the SQL, it is important for us to determine the atom number $N$ accurately because the SQL depends on $N$. We calibrate the atom number using the quantum shot noise of coherent states as detailed in ref. \cite{Luo2017TwinFock}. The inaccuracy of $N$ in our experiment is $7\%$ at $68\%$ confidence level, which gives a corresponding systematic error on sensitivity of 0.3\,dB. Thus the sensitivity enhancement we measured over the three-mode SQL is statistically credible.

\subsection{The distribution of the $\pi/2$-pulse-rotated prepared states in the Fock state basis}
Here, we discuss how to get the distribution of the $\pi/2$-pulse-rotated prepared states in the Fock state basis $\{|N_1, N_0, N_{-1}\rangle\}$, where $N_{m_F}$ denotes the number of atoms in the $m_F$ component, as shown in Figs. 3{\it C} and 3{\it D}
in the main text.

Ignoring atom loss, the state prepared in our experiment corresponds to a linear superposition of Dicke states $|l,0\rangle$ with different $l$, i.e.,
\begin{equation}\label{our}
 |\psi \rangle  = \sum\nolimits_{l = 0}^N {{d_l}|l,0\rangle },
\end{equation}
where the populated component with a smaller $l$ implies a higher excitation from the ground state.
For the state $|l,0\rangle$, a $\pi/2$ rotation along the $y$-axis transforms it into a superposition of $|l,m\rangle$ with different $m$, i.e.,
\begin{equation}
{e^{ - i\frac{\pi }{2}{\hat L_y}}}|l,0\rangle  = \sum\limits_{m =  - l}^l {d_{m,0}^l\left( {\pi /2} \right)|l,m\rangle } ,
\end{equation}
where the Wigner (small) d-matrix element $d_{m,0}^l\left( {\pi /2} \right) = \langle l,m|{e^{ - i\frac{\pi}{2}{\hat L_y}}}|l,0\rangle$ measures the projection amplitude onto the state ${| l,m\rangle }$ after $|l,0\rangle$ is rotated by $\pi$/2. Hence, after a $\pi/2$-pulse rotation along the $y$-axis, the superposition state $|\psi\rangle$ (Eq. (\ref{our})) becomes
\begin{equation}\label{psi_r}
|{\psi _{\pi/2}}\rangle  = {e^{ - i\frac{\pi }{2}{\hat L_y}}}\sum\limits_{l = 0}^N {{d_l}|l,0\rangle }  = \sum\limits_{m =  - N}^N {\sum\limits_{l = |m|}^N {{d_l}d_{m,0}^l\left( {\pi /2} \right)|l,m\rangle } }.
\end{equation}
Its distribution in the $L_z=m$ subspace reveals the interference of all the amplitudes of states $|l,m\rangle$ with $l\ge |m|$.
To obtain the distribution of state $|\psi_{\pi/2}\rangle$ in the Fock state basis, we first need the Fock-state representation for the state $|l,m\rangle$. Let us denote the expansion as
\begin{equation}\label{slm}
|l,m\rangle  = \sum\limits_{k = \max \left( {0, - m} \right)}^{{\rm{Floor}}\left[ {\left( {N - m} \right)/2} \right]} {f_k^{(l,m)}|k + m,N - 2k - m,k\rangle },
\end{equation}
with the projection coefficient $f_{k}^{(l,m)} \equiv \langle k + m,N - 2k - m,k|l,m\rangle $.

\subsubsection{The projection coefficient $f_{k}^{(l,m)}$}

\begin{figure}
\centering
  \includegraphics[width=1\columnwidth]{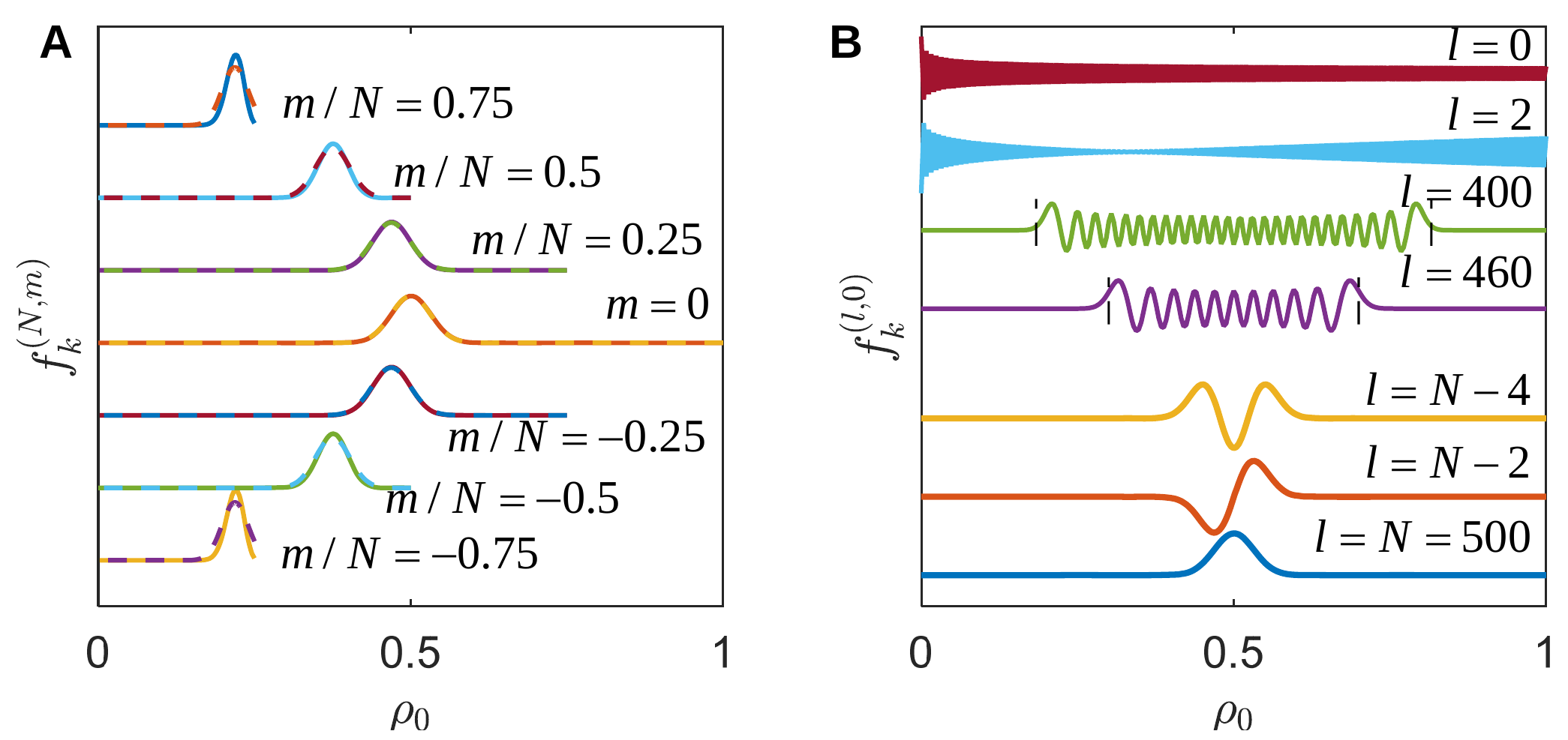}
  \caption{{Dicke state in the Fock state basis.} ({\it A}) The distributions for several Dicke states $|N,m\rangle$ in the Fock state basis. The solid lines are the exact results
from Eq. (\ref{Df}). The dashed lines denote the approximated Gaussian distributions given by Eq. (\ref{fapp}). ({\it B}) The distributions of a few selected Dicke states $|l,0\rangle$ in the Fock state basis. $N = 500$ is used for both figures.}\label{eigen}
\end{figure}

To obtain $f_{k}^{(l,m)}$, we can start from $m=l$, for which we have~\cite{Kawaguchi2012SpinorBEC}
\begin{equation}
|l,l\rangle  = {Z^{ - 1/2}}\hat{a}_1^{\dag l}{\hat{S}^{\dag \left( {N - l} \right)/2}}|0,0,0\rangle,
\end{equation}
where $\hat{S} = \left(\hat{a}_0^2 - 2{\hat{a}_1}{\hat{a}_{ - 1}}\right)/\sqrt{3}$ is the spin-singlet pair operator
and $Z$ is the normalization factor. Other states $|l,m\rangle$ with $m<l$ can be obtained by applying the spin lowering operator $\hat L_{-} = \sqrt{2} (\hat a_1 \hat a_0^{\dag} + \hat a_0 \hat a_{-1}^{\dag})$ repeatedly to $|l,l\rangle$ for $(l-m)$ times.
The expression for the projection coefficient $f_{k}^{(l,m)}$ in the $L_z=N_1-N_{-1}=m$
subspace is quite complicated~\cite{Wu1996PhysRevA.54.4534}. Some insight can nevertheless be gained from the simple case of $l=N$, for which we have
\begin{equation}\label{Df}
f_{k}^{(N,m)} = {2^{\left( {N - 2k - m} \right)/2}}\sqrt {C_N^kC_{N - k}^{k + m}{\rm{ /}}C_{2N}^{N + m}} .
\end{equation}
In the large $N$ limit and for small $m$, the above expression is well approximated by
\begin{equation}\label{fapp}
{f_{k}^{(N,m)}} \approx A \exp \left\{ { - {{\left( {k - {{\bar n}_{ - 1}}} \right)}^2}/{(2 \sigma ^2})} \right\},
\end{equation}
which is a Gaussian function centered
at ${{\bar n}_{ - 1}} = \frac{N}{4}{\left( {1 - \frac{m}{N}} \right)^2}$ with a width of $\sigma  = \sqrt {\frac{N}{8}\left[ {1 - {{\left( {\frac{m}{N}} \right)}^2}} \right]} $.
Here $A$ is the normalization factor so that $\sum\nolimits_k {{{| {f_{k}^{(N,m)}} |}^2}}  = 1$.
By using the properties of Gaussian function, it is easy to obtain the following quantities for the distribution of the Dicke state $|N,m\rangle$ in the Fock state basis,
\begin{eqnarray}\label{population}
\langle \rho_{-1} \rangle &=& \frac{{\langle {N_{ - 1}}\rangle }}{N} = \frac{{{{\bar n}_{ - 1}}}}{N} = \frac{1}{4}{\left( {1 - \frac{m}{N}} \right)^2}, \notag\\
\langle \rho_1 \rangle &=& \frac{{\langle {N_1}\rangle }}{N} = \frac{{{{\bar n}_{ - 1}} + m}}{N} = \frac{1}{4}{\left( {1 + \frac{m}{N}} \right)^2},\notag\\
\langle \rho_0 \rangle &=& \frac{{\langle {N_0}\rangle }}{N} = \frac{{N - \langle {n_1}\rangle  - \langle {n_{ - 1}}\rangle }}{N} = \frac{1}{2}\left[ {1 - {{\left( {\frac{m}{N}} \right)}^2}} \right], \notag\\
\Delta {\rho_{-1}} &=& \frac{\sigma}{\sqrt{2} N},\quad \Delta {\rho_1} = \frac{\sigma}{\sqrt{2} N}, \quad \Delta {\rho_0} = \frac{\sqrt{2}\sigma}{ N}.
\end{eqnarray}
For illustrations, we show the distributions of several Dicke states $|N,m\rangle$ with $N=500$ in the Fock state basis in Fig. \ref{eigen}{\it A}. The approximated Gaussian distributions (dashed lines) match well with the exact ones (solid lines) except when $m$ approaches $N$.

When calculating the projection coefficients $f_{k}^{(l,m)}$ for $l<N$, to avoid tedious bookkeeping of the expansion terms, we resort to numerical diagonalization of ${\bf \hat{L}}^2$, whose
eigenstate is $|l,m\rangle$. This is carried out by writing down the matrix form of ${\bf \hat{L}}^2\,[= 2N + (2\hat{N}_0-1)(N-\hat{N}_0) + (\hat{N}_1-\hat{N}_{-1})^2 + 2 \hat{a}_0^{\dag} \hat{a}_0^{\dag} \hat{a}_1 \hat{a}_{-1} + 2 \hat{a}_0 \hat{a}_0 \hat{a}_1^{\dag} \hat{a}_{-1}^{\dag}]$ in the Fock state basis within the $L_z=m$ subspace, $\{\left| {k + m,N - 2k - m,k} \right\rangle\}$, and then diagonalizing it numerically.

Fig. \ref{eigen}{\it B} shows the distributions of a few selected Dicke states $|l,0\rangle$ with a fixed $m=0$ but different $l$.
For $l \sim N$, the distribution takes a similar form to the spatial wave function for the $[(N-l)/2]$-th eigenstate of a one-dimentional harmonic oscillator~\cite{You2002PhysRevA.66.033611},
with its center located at $\langle \rho_0 \rangle \simeq 1/2$.
Two `classical turning points' separated by a distance of $\rho_{0(pp)}=\sqrt{2(1-l/N)}$ are clearly visible, which is larger for smaller $l$ (higher excitations). As in the case of $l=N$, the distributions for $|l,m\rangle$ with the same $l$ and different $m$ share similar shapes, with $m$-dependent centers and widths.
For $m \ll l$, we find $\langle \rho_0 \rangle \simeq [1 - {(m/N)^2}]/2$ and $\Delta {\rho _0} \simeq \sqrt {[1 - {{(l/N)}^2}][1 - {{(m/N)}^2}]/8}$, both decrease as $|m|$ increases.

\subsubsection{The rotated state $|\psi_{\pi/2}\rangle$ in the Fock state basis}
Having the projection structure of the state $|l,m\rangle$ in the Fock state basis at hand, it is now a simple exercise to find out the corresponding distribution of the state $|\psi_{\pi/2}\rangle$ (Eq. (\ref{psi_r})).
When expressed in the Fock state basis, the state $|\psi_{\pi/2}\rangle$ takes the following form
\begin{equation}
|{\psi _{\pi/2}}\rangle  = \sum\limits_{m =  - N}^N {\sum\limits_{k = \max \left( {0, - m} \right)}^{{\text {Floor}}\left[( {N - m} )/2\right]} {{\lambda _{m,k}}|k + m,N - 2k - m,k\rangle } } ,
\end{equation}
where the projection coefficient ${\lambda _{m,k}}$ in the $L_z=m$ subspace is given by
\begin{equation}\label{lambda}
{\lambda _{m,k}} = \sum\limits_{l = |m|}^N {{d_l}d_{m,0}^l\left( {\pi /2} \right)f_{k}^{(l,m)}},
\end{equation}
which is a weighted sum of all components $f_{k}^{(l,m)}$ with $l \ge |m|$.

For the numerical calculation of Eq. (\ref{lambda}), it is important to note that
when deriving $f_{k}^{(l,m)}$ from the diagonalization of $\hat {\bf L}^2$,
an overall negative sign can be added to the expression of an eigenstate $|l,m\rangle$ randomly. This `sign' problem can lead to erroneous results if not handled properly when considering superposition between the states $|l,m\rangle$ with different $l$. We fix this problem by enforcing an overall sign of $f_{k}^{(l,m)}$ for state $|l,m\rangle$ using the aforementioned property that the distribution for state $|l,m\rangle$ in the Fock state basis resembles a shifted one from that of the state $|l,0\rangle$. Given the expression of $f_{k}^{(l,0)}$ for state $|l,0\rangle$ (Fig. \,\ref{eigen}{\it B}), we can then fix the overall sign for all the other states $|l,m\rangle$.

As mentioned in the previous section, the width of the distribution for state $|l,m\rangle$ in the Fock state basis is larger for smaller $l$,
the distribution for the rotated superposition state $|\psi_{\pi/2}\rangle$ with components of various $l$ thus exhibits a wider spread in the Fock space compared with the rotated Dicke state $|N,0\rangle$, whose distribution is determined by $f_k^{(N,m)}$ in Eq. (\ref{Df}) with a singular contribution coming from $l=N$. The prepared state $|\psi\rangle$ (Eq. (\ref{our})) with a higher excitation (smaller $l$ populated)
displays a wider expansion after the $\pi/2$-pulse rotation. This observation is used by us to optimize the ramping parameters discussed in {\it Methods} by minimizing the fluctuation of $\rho_0$ for the final state.

\end{document}